\documentclass[usenatbib]{mn2e}
\usepackage{graphicx}
\usepackage{rotating}
\usepackage{txfonts}
\usepackage[british]{babel}
\usepackage{dcolumn}
\newcolumntype{d}{D{.}{.}{-1}}
\def\dhead#1{\multicolumn{1}{c}{#1}}
\def\twolines#1#2{$\kern-6pt\Big\{ {\textrm{#1\hfill}\atop\textrm{#2\hfill}}$}
\graphicspath{{.}{./FIGS/}}
\title[Milli-arcsecond properties of 10C sources in the Lockman Hole]{Milli-arcsecond properties of 10C sources in the Lockman Hole}

\label{firstpage}

\author[I.~H.~Whittam, J.~M.~Riley \& D.~A.~Green]{I.~H.~Whittam,\thanks{email:
\texttt{ihw24@mrao.cam.ac.uk}} J.~M.~Riley and D.~A.~Green\\
Astrophysics Group, Cavendish Laboratory, 19 J.~J.~Thomson Avenue,
Cambridge CB3 0HE}

\date{Accepted ---; received ---; in original form ---}

\pagerange{\pageref{firstpage}--\pageref{lastpage}}

\pubyear{2013}
\begin{document}

\maketitle

\begin{abstract}
We have used recent Very Long Baseline Interferometry (VLBI) observations by \citeauthor{2013A&A...551A..97M} with a resolution of $\approx 10$ mas to investigate the properties of faint sources selected from the Tenth Cambridge (10C) survey in the Lockman Hole. The 10C survey is complete to 0.5~mJy at 15.7~GHz and has a resolution of 30~arcsec. We have previously shown that this population is dominated by flat-spectrum sources below $\approx 1$~mJy, in disagreement with several models of the faint, high-frequency sky. We find that 33 out of the 51 10C sources in the VLBI field (65 percent) are detected by the VLBI observations. The sources detected by the VLBI observations must have a high brightness temperature, thus ruling out the possibility that this faint, high frequency population is dominated by starbursting or starforming sources and indicating that they must be Active Galactic Nuclei.
\end{abstract}


\section{Introduction}\label{section:intro}

In \citeauthor{2013MNRAS.429.2080W} (2013; hereafter Paper I) we studied a sample of 296 faint ($> 0.5$~mJy) sources selected from the Tenth Cambridge survey (10C; \citealt{2011MNRAS.415.2708D,2011MNRAS.415.2699F}) at 15.7~GHz in the Lockman Hole. The 10C survey was made with the Arcminute Microkelvin Imager (AMI; \citealt{2008MNRAS.391.1545Z}) and covers ten fields complete to 0.5~mJy, two of which are in the Lockman Hole. By matching the 10C catalogue to several lower-frequency surveys we investigated the radio spectral properties of the sources in this sample. We found a significant change in spectral index $\alpha$  with flux density $S$ (where $S \propto \nu^{-\alpha}$ for frequency $\nu$) -- the median spectral index between 15.7~GHz and 610~MHz changes from $\alpha = 0.75$ for 15.7-GHz flux densities $1.5 < S/ \rm mJy < 10$ to $\alpha = 0.08$ for lower flux densities ($0.3 < S/ \rm mJy < 0.8$) (full details of this study are given in Paper I). This suggests that a population of faint, flat-spectrum sources is emerging at flux densities $\lesssim 1 \textrm{ mJy}$.

The properties of the 10C sample were compared with those of a sample of sources selected from the SKADS Simulated Sky (S$^3$; \citealt{2008MNRAS.388.1335W,2010MNRAS.405..447W}) (full details are in Paper I).  We found that this simulation fails to reproduce the observed spectral index distribution and underpredicts the number of sources in the faintest flux density bin ($0.5 < S < 1~\rm mJy$). We discussed two possible origins for this discrepancy. The first is that the source population at this flux density level is not dominated by Active Galactic Nuclei (AGN) in the form of Fanaroff and Riley type I (FRI; \citealt{1974MNRAS.167P..31F}) sources, as predicted by the S$^3$ model, and is instead dominated by a population with flatter spectra, such as, for example, a population of starburst galaxies with unusually flat high-frequency spectra \citep{2013ApJ...768....2M}. The second possibility is that FRI sources are not modelled correctly in the simulation and in fact themselves have much flatter spectra at high frequencies. In Paper I we suggested that the first option is unlikely, as it would require the number of starburst sources to be incorrect in the simulation by at least a factor of ten.

Information about the structure of the sources in a sample can provide useful information about their nature, meaning that high resolution studies are valuable when investigating the properties of a population. The angular size of sources in the high frequency (20~GHz) source population at much higher flux densities ($S > 40~\rm mJy$) has recently been investigated by \citet{2013MNRAS.tmp.1815C}. They used data from a 6~km baseline to split the sample into different populations and investigate their properties. They found that 77 percent of their sample are compact AGNs and 23 percent are extended AGN-powered sources. Here we study a sample of high frequency sources which are considerably fainter on even smaller angular scales.

In this paper we use recently published VLBI data (\citealt{2013A&A...551A..97M}, hereafter M2013) to investigate further the nature of this faint 15.7-GHz source population. M2013 have recently conducted wide-field observations with the Very Long Baseline Array (VLBA) at 1.4~GHz of part of the Lockman Hole. The VLBA provides milliarcsec-scale resolution, so if a source is detected the emission must come from a very compact region and the brightness temperature must be very high ($\gtrsim 10^6$ K). Any stellar non-thermal sources, such as supernova remnants, at redshift $\gtrsim 0.1$ cannot have a surface brightness this high and thus any objects with $z \gtrsim 0.1$ which are detected must originate in AGN.

In Section \ref{section:methods} the different samples used in this paper are defined and the methods used to investigate the properties of these samples are described. The results and discussion are split into two parts -- the first part, Section \ref{section:results1}, discusses the properties of the 10C sources in the VLBA survey area and the second part, Section \ref{section:results2}, contains further insights into the nature of the M2013 VLBA sources. The conclusions are given in Section \ref{section:conclusions}.

Throughout this paper the term `flat spectrum' refers to an object with spectral index $\alpha \leqslant 0.5$, `steep spectrum' to an object with $\alpha > 0.5$ and `rising' to an object with $\alpha < 0$.

\section{Sample definition and properties}\label{section:methods}

\subsection{\citeauthor{2013A&A...551A..97M} observations}

M2013 used a 1.4-GHz VLA image of the Lockman Hole field by \citet{2009MNRAS.397..281I} to identify a sample of sources which could in principle be detected with the VLBA observations provided all the flux comes from a very compact (milli-arcsecond scale) region. They defined a source as `detectable' if its peak VLA flux density is $\gtrsim~6$ times the noise in the VLBA image at that point. They found that 217 sources fitted these criteria -- we refer to this sample as the Middelberg sample.

M2013 made naturally weighted images, with a median resolution of $11.9 \times 9.4~\rm{mas}^2$, at the positions of each of the 217 sources to detect, or place limits on, any millarcsecond-scale components. Images with uniform weighting, with a median resolution of $7.4 \times 5.5~\rm{mas}^2$, were used to calculate the integrated flux density of each detected source. Sixty-five of the sources in the Middelberg sample were actually detected in the VLBA observations, using the $6~\sigma$ threshold for detection. These sources form the `VLBA-detected' sample.

Redshift information is available for 47 out of the 65 VLBA-detected sources from the \citet{2012ApJS..198....1F} photometric redshift catalogue. The redshift values for these sources all lie in the range $0.2 < z < 4.2$, meaning that all VLBA-detected sources must have the very high brightness temperatures which can only be found in AGN (see Section \ref{section:intro}).

\subsection{Data used and sample definition}\label{section:samples}

\begin{figure}
\centerline{\includegraphics[width=8cm,angle=270,clip=,bb=53 145 567 697]{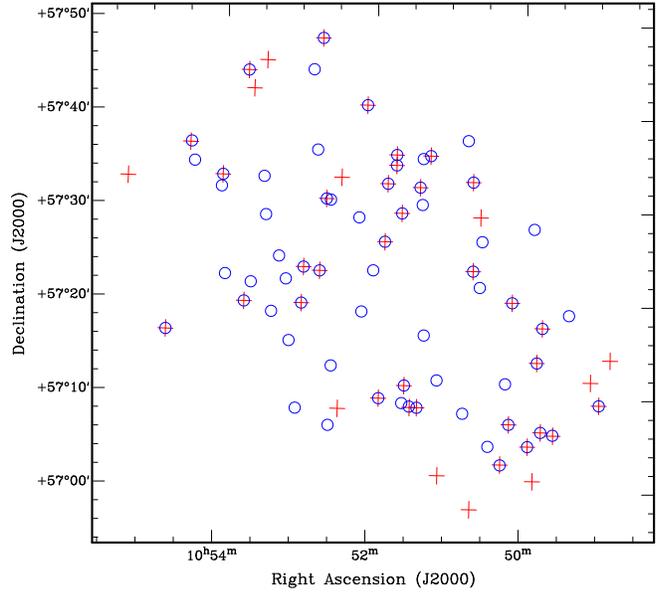}}
\caption{Positions of 10C and VLBI sources in the Lockman Hole. Red crosses show 10C sources in the Middelberg sample (i.e. the 10C sources which should, in theory, be detectable by the VLBA observations if they are compact). Blue circles show the 65 Middelberg sources which are detected by the VLBA.}\label{fig:source-map}
\end{figure}

The 10C survey was observed with AMI at 15.7~GHz and covers ten fields, one of which contains the area covered by the M2013 VLBA survey in the Lockman Hole. In this area the 10C survey is complete to 0.5~mJy. In Paper I we presented spectral information about the 10C sources obtained by matching the catalogue to a deep Giant Metrewave Radio Telescope (GMRT) survey at 610~MHz \citep{2008MNRAS.387.1037G,2010BASI...38..103G}, a Westerbork Synthesis Radio Telescope (WSRT) survey at 1.4 GHz \citep{2012rsri.confE..22G}, the National Radio Astronomy Observatory (NRAO) Very Large Array (VLA) Sky Survey (NVSS; \citealt{1998AJ....115.1693C}) and the Faint Images of the Radio Sky at Twenty-one centimetres (FIRST; \citealt{1997ApJ...475..479W}). The M2013 observations consist of three pointings with an r.m.s. noise of 24~$\muup$Jy beam$^{-1}$ towards the pointing centre. 

The 10C and Middelberg catalogues were matched in Topcat\footnote{See http://www.starlink.ac.uk/topcat/} using a match radius of 15 arcsec. This match radius was chosen to minimise the number of false detections by matching to a shifted distribution of sources in the same way as described in Paper I. There are 51 10C sources in the VLBA survey area, of which 44 match to a source in the Middelberg sample. These 44 sources should therefore be bright enough at 1.4-GHz to be detectable with the VLBA if all their flux is in a compact core. Thirty-three of these 44 10C sources are actually detected by the VLBA observations. 

The seven 10C sources which are not in the Middelberg sample (which would therefore not be detectable by the VLBA observations) have rising spectra, with typical values of $\alpha \sim -0.5$, and therefore fall below the VLBA-detectable flux density limit at 1.4~GHz (note that all 10C sources are found in the \citet{2009MNRAS.397..281I} deep VLA image).

In two separate cases a second VLBA source falls within the contours of a 10C source which is already matched to a different VLBA source. In both cases, the second VLBA source is more than 15 arcsec (the radius used for matching) away from the 10C source position, so is not counted as a match. The first case (10C ID 10CJ105115+573552) appears as an extended source in the 10C map (resolution $\approx 30$~arcsec). In the higher-resolution GMRT map (synthesised beam $5 \times 6~\rm arcsec$) this source is resolved into two separate compact sources and the two VLBA sources correspond to these two separate sources. It is likely that both VLBA sources are seen in the 10C map but they are blended together. If this is the case, it will mean that the 15.7-GHz flux density of the matched VLBA source is over estimated, as it also contains some contribution from the second VLBA source. The matched VLBA source has a steep spectum ($\alpha^{15.7}_{1.4} = 0.65$), so if the 15.7-GHz flux density is over-estimated the spectrum may be steeper still. In the second case (10C ID 10CJ105237+573058), the higher-resolution GMRT images reveal that the VLBA source which is closer to the 10C source position corresponds to the core of an extended double-lobed radio source, while nothing is visible in the GMRT map at the position of the second VLBA source. This is probably because this source is very faint ($S_{1.4~\rm GHz} = 0.14~\rm mJy$), so falls below the detection limit of the GMRT observations.

The positions of the 10C sources in the Middelberg sample and the VLBA-detected sources are shown in Fig. \ref{fig:source-map} and the different samples are summarised in Table \ref{tab:samples}.

\begin{table}
\begin{center}
\begin{minipage}{7cm}
\caption{Summary of the different samples used in this paper.}\label{tab:samples}
\begin{tabular}{ll}\hline
Description                         & Number of sources \\\hline
Middelberg VLBA-detectable sources  & 217\\
Middelberg sources in 10C sample    & 44\\
Middelberg sources detected by VLBA & 65\\
10C sources detected by VLBA        & 33\\\hline
\end{tabular}
\end{minipage}
\end{center}
\end{table}

\subsection{VLBA sources without a match in 10C}\label{section:unmatched-VLBI}

Thirty of the Middelberg sources which are detected by the VLBA observations do not have a match in the 10C catalogue. For each of these unmatched VLBA sources the 10C image was examined to see if there was a source present which was below the 10C catalogue limit. In 14 cases a source was visible within 15 arcsec of the VLBA position  with a peak flux density greater than $3 \sigma$. For these 14 sources the 15.7-GHz flux density was estimated from the 10C image using the AIPS\footnote{Astronomical Image Processing System, see http://www.aips.nrao.edu/.} task JMFIT. In four cases JMFIT did not converge due to the presence of a nearby bright source; for these sources the flux density was found manually using the AIPS task TVSTAT. For the remaining unmatched sources an upper limit of three times the local noise in the 10C map was placed on the 15.7-GHz flux density.

\subsection{Spectral index}\label{section:alpha}

 In sections \ref{section:results1} and \ref{section:results2} radio spectral index, $\alpha$, (where $S = \nu ^{-\alpha}$ for flux density $S$ at frequency $\nu$), is used to investigate source properties. 

In Section \ref{section:results1}, for sources in the 10C sample, $\alpha^{15.7}_{1.4}$ and $\alpha^{15.7}_{610}$ values from Paper I are used.

In Section \ref{section:results2} the spectral index, or a lower limit on the spectral index, between 15.7~GHz and 1.4~GHz ($\alpha^{15.7}_{1.4}$) was calculated for all 65 sources detected by the VLBA observations. For consistency, the 1.4-GHz flux densities used were the integrated flux densities from the VLA observations by \citet{2009MNRAS.397..281I} used in M2013. The 15.7-GHz flux densities were either the integrated flux densities taken from the 10C catalogue or, for those sources without a match in the 10C catalogue, the values determined from the 10C map or the upper limit on the 10C flux density derived from the 10C map (see Section \ref{section:unmatched-VLBI}).

\subsection{Morphology}\label{section:morph-methods}

In this study we use two different measures of source morphology, which give us information about the size of the source on different scales, as follows.

(1) VLA size, $\theta_{\rm{VLA}}$ -- the maximum angular size taken from the \citet{2009MNRAS.397..281I} 1.4-GHz VLA catalogue. For single sources this value corresponds to twice the maximum Full Width Half Maximum (FWHM) of the fitted gaussian and gives information about whether or not the source is resolved on an arcsecond scale. The FWHM synthesised beam size is $\approx 4$~arcsec so values of $\theta_{\rm VLA} \gtrsim 9$~arcsec indicate that the source is resolved.

(2) Ratio $R$ of the VLA integrated flux density to the VLBA integrated flux density ($R = S_{\rm VLA}/S_{\rm VLBA}$). This gives information about whether the source is dominated by compact emission ($\lesssim 10~\rm{mas}$), in which case $R \approx 1$, or whether there is significant emission from a more extended region.

\section{Properties of the 10C sources}\label{section:results1}

\def\vpad{{\Large$\mathstrut$}}

%
%

\makeatletter
\let\@makecaption=\SFB@makefigurecaption
\makeatother
\setlength{\rotFPtop}{0pt plus 1fil}
\setlength{\rotFPbot}{0pt plus 1fil}

\begin{sidewaystable*}
\caption{Properties of 10C sources in the Middelberg sample (which should be detectable by the VLBA if they are compact). Sources which are detected by the VLBA observations are marked with a $\ast$.
All flux densities listed are integrated flux densities. All spectral index values are from Paper~I.}\label{tab:results_M10C}
\begin{center}
\fontsize{5.9}{8.0}\selectfont
\rm
\begin{tabular}{r@{}lllddddcddcddldddrrrrl}
\hline

\vpad & 10C ID$^{a}$ & RA (J2000) & Dec (J2000) & \dhead{$S_{\rm 10C}$} & \dhead{$\sigma\_S_{\rm 10C}$} & \dhead{$\alpha^{15.7}_{0.61}$} & \dhead{$\sigma\_\alpha^{15.7}_{0.61}$} & Flag1$^{\rm b}$ & \dhead{$\alpha^{15.7}_{1.4}$} & \dhead{$\sigma\_\alpha^{15.7}_{1.4}$}  & Flag2$^{\rm c}$ & \dhead{$\alpha^{1.4}_{0.61}$} & \dhead{$\sigma\_\alpha^{1.4}_{0.61}$} & Name$^{\rm d}$ & \dhead{$S_{\rm VLA}$} & \dhead{$\sigma\_S_{\rm VLA}$} & \dhead{$S_{\rm VLBA}$} & \dhead{$\sigma\_S_{\rm VLBA}$} & \dhead{$\theta_{\rm{VLA}}$$^{\rm e}$} & \dhead{$R^{\rm f}$} \\
\vpad & & & & \dhead{/mJy}  & \dhead{/mJy} & & & & & & & & & & \dhead{/mJy} & \dhead{/mJy} & \dhead{/mJy} & \dhead{/mJy} & \dhead{/arcsec} &  \\\hline
\vpad
        & 10CJ104849+571417 & 10:48:49.47 & +57:14:17.48 & 1.82  & 0.17  & 0.74  &  0.03  & 0 & 0.71  & 0.04  & 1 & 0.8   & 0.06 & J104849.8+571415  &  10.84   & 0.048 &        &        & 9.0   & \\
$\ast$  & 10CJ104858+570933 & 10:48:58.99 & +57:09:33.03 & 0.50  & 0.16  & 0.50  &  0.14  & 0 & 0.47  & 0.13  & 2 & 0.59  & 0.39 & J104858.3+570925  &  1.384   & 0.046 & 0.642  & 0.121  & 8.8   & 2.16\\
        & 10CJ104906+571156 & 10:49:06.21 & +57:11:56.35 & 1.86  & 0.21  & 0.69  &  0.04  & 0 & 0.83  & 0.05  & 1 & 0.29  & 0.06 & J104905.1+571151  &  14.612  & 0.730 &        &        & 37.9  & \\
$\ast$  & 10CJ104934+570613 & 10:49:34.42 & +57:06:13.69 & 2.09  & 0.17  & 0.76  &  0.03  & 0 & 0.78  & 0.04  & 1 & 0.72  & 0.05 & J104934.3+570608  &  14.787  & 0.082 & 0.295  & 0.080  & 19.6  & 50.13\\
$\ast$  & 10CJ104943+571739 & 10:49:43.82 & +57:17:39.14 & 1.17  & 0.13  & -0.16 &        & 1 & -0.08 & 0.05  & 2 & -0.39 &      & J104943.8+571737  &  1.013   & 0.025 & 0.666  & 0.084  & 8.9   & 1.52\\
$\ast$  & 10CJ104944+570635 & 10:49:44.12 & +57:06:35.92 & 0.81  & 0.16  & 0.08  &  0.07  & 0 & 0.30  & 0.08  & 1 & -0.54 & 0.11 & J104944.1+570628  &  1.275   & 0.034 & 0.916  & 0.113  & 9.1   & 1.39\\
$\ast$  & 10CJ104947+571355 & 10:49:47.49 & +57:13:55.24 & 0.68  & 0.12  & 0.07  &        & 1 & -0.02 & 0.07  & 2 & 0.32  &      & J104947.8+571354  &  0.607   & 0.026 & 0.490  & 0.073  & 9.0   &  1.24\\
        & 10CJ104950+570117 & 10:49:50.50 & +57:01:17.79 & 0.66  & 0.12  & 0.70  &  0.06  & 0 & 0.86  & 0.08  & 1 & 0.22  & 0.13 & J104950.4+570120c &  1.047   & 0.083 &        &        & 28.1  & \\
$\ast$  & 10CJ104954+570456 & 10:49:54.30 & +57:04:56.77 & 5.64  & 0.32  & -0.63 &  0.04  & 0 & -0.71 & 0.03  & 2 & -0.42 & 0.14 & J104954.2+570456  &  1.376   & 0.034 & 0.633  & 0.091  & 8.9   & 2.17\\
$\ast$  & 10CJ105007+572020 & 10:50:07.93 & +57:20:20.28 & 1.18  & 0.10  & 0.11  &  0.04  & 0 & 0.28  & 0.04  & 2 & -0.38 & 0.09 & J105008.1+572018  &  1.668   & 0.019 & 1.195  & 0.127  & 9.9   & 1.70\\
$\ast$  & 10CJ105009+570724 & 10:50:09.71 & +57:07:24.56 & 1.05  & 0.18  & 0.57  &  0.05  & 0 & 0.67  & 0.08  & 1 & 0.29  & 0.12 & J105010.4+570724b &  2.302   & 0.023 & 2.335  & 0.240  & 8.6   & 2.40\\
$\ast$  & 10CJ105015+570258 & 10:50:15.38 & +57:02:58.44 & 0.51  & 0.13  & 0.68  &  0.08  & 0 & 0.77  & 0.11  & 1 & 0.39  & 0.15 & J105015.6+570258  &  3.155   & 0.045 & 0.983  & 0.117  & 13.9  & 3.21\\
        & 10CJ105034+572922 & 10:50:34.05 & +57:29:22.91 & 1.20  & 0.12  & -0.01 &  0.04  & 0 & -0.52 & 0.04  & 2 & 1.46  & 0.12 & J105034.2+572922  &  0.336   & 0.019 &        &        & 9.3   & \\
        & 10CJ105038+565810 & 10:50:38.75 & +56:58:10.04 & 0.54  & 0.14  & 0.43  &  0.09  & 0 & 0.48  & 0.11  & 2 & 0.30  & 0.10 & J105039.1+565806  &  1.520   & 0.043 &        &        & 9.9   & \\
$\ast$  & 10CJ105039+572339 & 10:50:39.56 & +57:23:39.77 & 1.08  & 0.12  & 0.34  &  0.04  & 0 & 0.76  & 0.08  & 1 & -0.88 & 0.19 & J105039.6+572336  &  4.936   & 0.022 & 3.972  & 0.399  & 13.1  & 1.24\\
$\ast$  & 10CJ105040+573308 & 10:50:40.65 & +57:33:08.68 & 0.89  & 0.12  & 0.16  &        & 1 & 0.00  & 0.06  & 2 & 0.61  &      & J105040.7+573308  &  0.703   & 0.019 & 0.984  & 0.112  & 8.7   & 0.71\\
        & 10CJ105104+570148 & 10:51:04.45 & +57:01:48.35 & 0.54  & 0.15  & 0.64  &  0.09  & 0 & 0.84  & 0.13  & 1 & 0.06  & 0.16 & J105104.7+570150b &  1.409   & 0.051 &        &        & 14.5  & \\
$\ast$  & 10CJ105115+573552 & 10:51:15.54 & +57:35:52.32 & 0.53  & 0.14  & 0.72  &  0.08  & 0 & 0.65  & 0.11  & 1 & 0.92  & 0.03 & J105115.0+573552  &  2.362   & 0.020 & 0.133  & 0.049  & 9.1   & 17.76\\
$\ast$  & 10CJ105122+570854 & 10:51:22.18 & +57:08:54.76 & 1.26  & 0.12  & 0.84  &  0.03  & 0 & 0.87  & 0.05  & 1 & 0.73  & 0.06 & J105122.1+570854  &  10.644  & 0.532 & 10.562 & 0.921  & 25.8  & 1.01\\
$\ast$  & 10CJ105123+573229 & 10:51:23.11 & +57:32:29.40 & 0.38  & 0.10  & 0.49  &  0.08  & 0 & 0.78  & 0.13  & 1 & -0.38 & 0.21 & J105122.9+573228  &  1.434   & 0.016 & 1.343  & 0.138  & 8.9   & 1.07\\
$\ast$  & 10CJ105128+570901 & 10:51:28.10 & +57:09:01.68 & 2.25  & 0.16  & 0.67  &  0.02  & 0 & 0.70  & 0.03  & 1 & 0.56  & 0.05 & J105127.8+570854  &  12.570  & 0.628 & 0.506  & 0.062  & 29.4  & 24.84\\
$\ast$  & 10CJ105132+571114 & 10:51:32.63 & +57:11:14.59 & 2.92  & 0.19  & 0.34  &  0.03  & 0 & 0.60  & 0.04  & 1 & -0.42 & 0.13 & J105132.4+571114  &  12.756  & 0.637 & 1.157  & 0.120  & 158.7 & 11.03\\
$\ast$  & 10CJ105136+572944 & 10:51:36.99 & +57:29:44.40 & 1.12  & 0.10  & 0.41  &  0.03  & 0 & 0.38  & 0.08  & 1 & 0.50  & 0.22 & J105137.0+572940  &  2.532   & 0.012 & 0.402  & 0.049  & 10.2  & 6.30\\
$\ast$  & 10CJ105142+573447 & 10:51:42.02 & +57:34:47.81 & 0.79  & 0.10  & -0.04 &        & 1 & 0.02  & 0.05  & 2 & -0.21 &      & J105142.0+573447  &  1.090   & 0.015 & 1.168  & 0.122  & 8.9   & 0.93\\
$\ast$  & 10CJ105142+573557 & 10:51:42.07 & +57:35:57.97 & 1.95  & 0.15  & 0.45  &  0.02  & 0 & 0.37  & 0.05  & 1 & 0.67  & 0.10 & J105142.1+573554  &  1.268   & 0.017 & 1.243  & 0.130  & 11.8  & 3.97\\
$\ast$  & 10CJ105148+573245 & 10:51:48.76 & +57:32:45.04 & 0.52  & 0.10  & 0.08  &        & 1 & 0.24  & 0.08  & 2 & -0.41 &      & J105148.7+573248  &  0.807   & 0.018 & 0.944  & 0.099  & 8.6   & 0.85\\
$\ast$  & 10CJ105150+572637 & 10:51:50.02 & +57:26:37.10 & 0.30  & 0.09  & 0.28  &        & 1 & 0.09  & 0.12  & 2 & 0.82  &      & J105150.1+572635  &  0.118   & 0.009 & 0.272  & 0.036  & 9.2   & 0.43\\
$\ast$  & 10CJ105152+570950 & 10:51:52.33 & +57:09:50.63 & 0.37  & 0.11  & 0.81  &  0.09  & 0 & 0.85  & 0.14  & 1 & 0.70  & 0.17 & J105152.4+570950  &  3.118   & 0.019 & 2.965  & 0.216  & 8.8   & 1.05\\
$\ast$  & 10CJ105206+574111 & 10:52:06.40 & +57:41:11.78 & 3.35  & 0.20  & 0.46  &  0.02  & 0 & 0.50  & 0.03  & 1 & 0.35  & 0.06 & J105206.5+574109  &  10.542  & 0.021 & 8.420  & 0.555  & 8.7   & 1.25\\
        & 10CJ105224+570837 & 10:52:24.96 & +57:08:37.66 & 0.42  & 0.12  & 0.53  &  0.09  & 0 & 0.95  & 0.14  & 1 & -0.69 & 0.19 & J105224.5+570838  &  2.157   & 0.028 &        &        & 15.4  & \\
        & 10CJ105225+573323 & 10:52:25.84 & +57:33:23.31 & 0.58  & 0.15  & 0.85  &  0.08  & 0 & 0.90  & 0.11  & 1 & 0.69  & 0.10 & J105225.6+573322  &  4.873   & 0.024 &        &        & 13.1  & \\
$\ast$  & 10CJ105237+573058 & 10:52:37.01 & +57:30:58.85 & 5.16  & 0.30  & 1.06  &  0.02  & 0 & 1.05  & 0.03  & 1 & 1.09  & 0.04 & J105237.3+573103  &  63.323  & 3.166 & 0.607  & 0.066  & 48.4  & 104.32\\
$\ast$  & 10CJ105240+572322 & 10:52:40.87 & +57:23:22.96 & 0.54  & 0.09  & 0.42  &  0.06  & 0 & 0.50  & 0.07  & 1 & 0.19  & 0.07 & J105241.4+572320  &  1.740   & 0.012 & 0.184  & 0.030  & 9.9   & 9.46\\
$\ast$  & 10CJ105243+574817 & 10:52:43.34 & +57:48:17.11 & 1.00  & 0.13  & -0.01 &  0.05  & 0 & 0.26  & 0.05  & 2 & -0.80 & 0.13 & J105243.3+574813  &  1.857   & 0.039 & 1.237  & 0.159  & 9.0   & 1.50\\
$\ast$  & 10CJ105253+572348 & 10:52:53.86 & +57:23:48.65 & 0.47  & 0.08  & 0.11  &        & 1 & -0.36 & 0.07  & 2 & 1.46  &      & J105254.2+572341  &  0.157   & 0.010 & 0.140  & 0.028  & 8.9   & 1.12\\
$\ast$  & 10CJ105255+571949 & 10:52:55.07 & +57:19:49.35 & 0.39  & 0.11  & 0.75  &  0.09  & 0 & 0.91  & 0.13  & 1 & 0.28  & 0.18 & J105255.3+571950  &  3.107   & 0.013 & 2.921  & 0.293  & 8.8   & 1.06\\
        & 10CJ105327+574546 & 10:53:27.36 & +57:45:46.09 & 0.81  & 0.12  & 0.73  &  0.05  & 0 & 0.89  & 0.07  & 1 & 0.28  & 0.09 & J105327.6+574543  &  6.618   & 0.330 &        &        & 30.7  & \\
        & 10CJ105337+574242 & 10:53:37.29 & +57:42:42.26 & 0.38  & 0.09  & 0.26  &  0.08  & 0 & 0.70  & 0.10  & 2 & -1.02 & 0.14 & J105337.3+574240  &  0.995   & 0.026 &        &        & 10.5  & \\
$\ast$  & 10CJ105341+571951 & 10:53:41.01 & +57:19:51.29 & 0.55  & 0.09  & 0.47  &  0.06  & 0 & 0.42  & 0.07  & 2 & 0.60  & 0.06 & J105340.9+571952  &  1.554   & 0.015 & 0.283  & 0.044  & 8.9   & 5.49\\
$\ast$  & 10CJ105342+574438 & 10:53:42.19 & +57:44:38.11 & 1.80  & 0.14  & 0.50  &  0.03  & 0 & 0.59  & 0.04  & 1 & 0.23  & 0.09 & J105342.1+574436  &  7.825   & 0.391 & 2.152  & 0.194  & 40.4  & 3.64\\
$\ast$  & 10CJ105400+573324 & 10:54:00.97 & +57:33:24.63 & 0.67  & 0.13  & 0.55  &  0.06  & 0 & 0.62  & 0.10  & 1 & 0.37  & 0.17 & J105400.5+573321  &  3.016   & 0.025 & 0.222  & 0.043  & 14.8  & 13.59\\
$\ast$  & 10CJ105425+573700 & 10:54:25.05 & +57:37:00.78 & 25.47 & 0.92  & 0.79  &  0.01  & 0 & 0.83  & 0.02  & 1 & 0.65  & 0.01 & J105427.0+573644  &  181.160 & 9.058 & 1.658  & 0.176  & 67.6  & 109.26\\
$\ast$  & 10CJ105441+571640 & 10:54:41.45 & +57:16:40.84 & 0.78  & 0.10  & 0.67  &  0.04  & 0 & 0.63  & 0.07  & 1 & 0.78  & 0.14 & J105442.1+571639  &  2.778   & 0.037 & 0.945  & 0.136  & 8.9   & 2.94\\
        & 10CJ105515+573256 & 10:55:15.50 & +57:32:56.87 & 0.63  & 0.10  & 0.89  &  0.05  & 0 & 0.94  & 0.07  & 1 & 0.75  & 0.08 & J105516.0+573257  &  7.104   & 0.355 &        &        & 29.6  & \\
\hline\end{tabular}								
\end{center}									
								
Notes:\\										
a) ID and positions from the 10C catalogue.\\					
b) Flag1 = 1 if value of $\alpha^{15.7}_{0.61}$ is an upper limit, Flag1 = 0 otherwise.\\
c) Indicates 1.4-GHz flux density used to calculate $\alpha^{15.7}_{1.4}$. Flag2 =  1 if value is from NVSS, Flag2 = 2 if value is from \citet{2012rsri.confE..22G} WSRT map.\\
d) Source name from M2013 and \citeauthor{2009MNRAS.397..281I} catalogues.\\
e) Maximum angular size taken from the \citet{2009MNRAS.397..281I} 1.4-GHz VLA catalogue. For single sources this value corresponds to twice the maximum Full Width Half Maximum (FWHM) of the synthesised beam (see Section \ref{section:morph-methods}).\\
f) Ratio of the VLA integrated flux density to the VLBA integrated flux density (see Section \ref{section:morph-methods}).\\
\end{sidewaystable*}			    

\subsection{VLBA detections}\label{section:detections}

\begin{figure}
\centerline{\includegraphics[width=8cm]{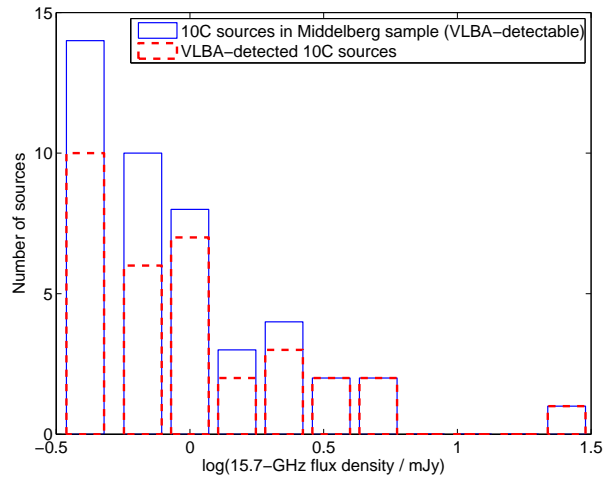}}
\caption{15.7-GHz flux density distribution for the 44 10C sources in the Middelberg sample, showing the 33 sources which are detected by the VLBA observations separately.}\label{fig:flux}
\end{figure}

\begin{figure}
\centerline{\includegraphics[width=8cm]{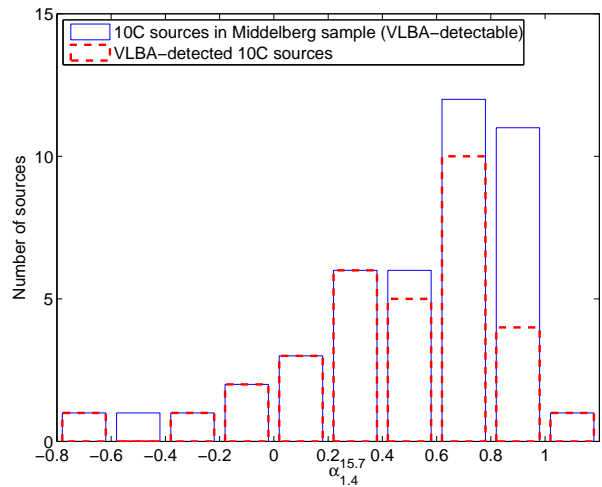}}
\caption{Spectral index distribution for the 44 10C sources in the Middelberg sample, showing the 33 sources which are detected by the VLBA observations separately. Spectral index values are taken from Paper I.}\label{fig:alpha}
\end{figure}

The spectral index and morphology properties of the 44 10C sources in the Middelberg sample are summarised in Table \ref{tab:results_M10C}. Thirty-three out the 44 sources (75 percent) in this sample are detected by the VLBA observations. This is much higher than the percentage of sources in the total Middelberg sample which are detected by the VLBA observations (65 out of 217 i.e. 30 percent). However nothing significant can be deduced from this since the sources in the 10C sample have higher 1.4-GHz flux densities than the majority of sources in the Middelberg sample (e.g. 14 percent of the 10C sources in the Middelberg sample have $S_{1.4 \rm{GHz}} < 1$~mJy compared with 75 percent of sources in the total Middelberg sample) and, as shown in M2013, the probability of detection is a strong function of flux density.

The 15.7-GHz flux density distributions of the 44 10C sources in the Middelberg sample and the 33 10C sources which are detected in the VLBA observations are shown in Fig \ref{fig:flux}. The flux density distribution of the 10C VLBA-detected sources is similar to that of the 10C sources in the Middelberg sample (although, significantly, all seven of the brightest sources with $S_{15.7~\rm{GHz}} > 2$~mJy are detected by the VLBA observations). There are VLBA-detected sources with 15.7-GHz flux densities as low as $S_{15.7~\rm{GHz}} = 0.3$~mJy so we are finding VLBI components in the majority of the faintest 10C sources.

The fact that 75 percent of the 10C sources in the Middelberg sample are detected by the VLBA observations, and therefore must be AGN, rules out the possibility discussed in Paper I that the 10C population is dominated by starburst galaxies. These results therefore support the conclusions in Paper I that the majority of the 10C population are AGN, such as FRI sources. Thus the proportions of sources in S$^3$, which predicts that the faint, high frequency population is dominated by FRI sources, may well be correct. If this is the case, the emission from the FRI sources must be modelled incorrectly, because, as described in Section \ref{section:intro}, the simulation does not accurately reproduce the observed spectral index distribution. This is probably due to incorrect modelling of the emission from the cores of FRI sources (for example, there is no attempt to include the effectes self-absorption in compact objects), as well as the lack of spectral aging in the overall source model.

\subsection{Spectral properties}

The spectral index distribution of the 44 10C sources in the Middelberg sample and the 33 detected in the VLBA observations are shown in Fig. \ref{fig:alpha}, using the values from Paper I. The sources which are not detected in the VLBA observations tend to have steeper spectra than those which are detected, with 9 of the 11 undetected sources having $\alpha^{15.7}_{1.4} > 0.5$ (of the other two, one has $\alpha^{15.7}_{1.4} = 0.47$ and the other has a rising spectrum and is discussed below). However, many of the sources which are detected by the VLBA observations also have steep spectra. In fact, out of the 26 steep spectrum 10C sources, 17 (65 percent) are detected by the VLBA observations.

\begin{figure}
\centerline{\includegraphics[width=8cm]{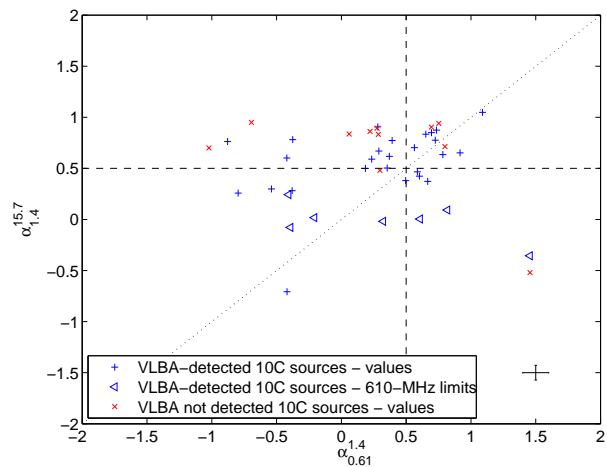}}
\caption{Colour-colour plot for 10C sources in the Middelberg sample, showing those which are and are not detected by the VLBA observations separately. Spectral index values for all sources are taken from Paper I. Sources which have an upper limit at 610~MHz are shown by triangles and could move to the left. The horizontal and vertical dashed lines are at $\alpha = 0.5$, the cutoff between steep and flat spectrum sources. The diagonal dotted line is at $\alpha^{15.7}_{1.4} = \alpha^{1.4}_{0.61}$, sources which lie above and to the left of this line have spectra which are steeper between 15.7 and 1.4~GHz than between 1.4~GHz and 610~MHz. For clarity individual error bars have been omitted but a point representing the median errors is shown in the bottom right hand corner for reference. }\label{fig:c-c}
\end{figure}

Further insights into the spectral shape of 10C sources in the Middelberg sample can be gained from Fig. \ref{fig:c-c}, a radio colour--colour plot which uses the spectral index values $\alpha^{15.7}_{1.4}$ and $\alpha^{1.4}_{0.61}$ from Paper I. The majority (9 out of 11) of the sources which are not detected by the VLBA observations (shown by crosses, red in the online version) have $\alpha^{15.7}_{1.4} > \alpha^{1.4}_{0.61}$ and therefore have spectra which steepen at higher frequencies.  There is a larger spread in values of $\alpha^{1.4}_{0.61}$ than of $\alpha^{15.7}_{1.4}$, particularly towards negative values; 14 of the sources which display a flat or rising spectrum between 610~MHz and 1.4~GHz ($\alpha^{1.4}_{0.61} < 0.5$) have turned over and have a steep spectrum ($\alpha > 0.5$) by 15.7~GHz.

\begin{figure}
\centerline{\includegraphics[width=8cm]{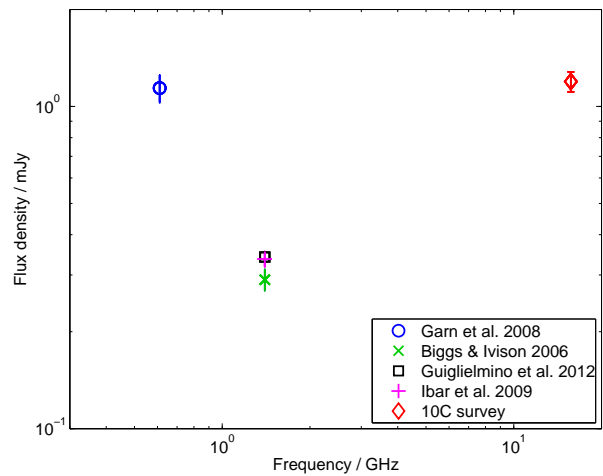}}
\caption{Spectrum for source 10CJ105034+572922 which has an unusual spectral shape and is not detected by the VLBA observations. Where error bars are not shown they are smaller than the symbol plotted.}\label{fig:spectrum}
\end{figure}

All but two of the 10C sources detected by the VLBA observations have flat spectra; this is not suprising as flat spectrum emission generally originates from compact core regions, which are much more likely to have a high enough surface brightness to be visible in the VLBA observations. Conversely, steep spectrum emission tends to come from more extended regions, which do not have high enough surface brightness to be detectable with the VLBA observations -- consistent with the fact that most of the 10C sources not detected by the VLBA observations have steep spectra. The majority of these undetected sources have a spectrum which steepens with frequency, supporting the idea that these sources do not have a dominant core, as this would contribute more to the flux density at higher frequencies.

As mentioned above, one 10C source not detected by the VLBA observations has a rising spectrum, with $\alpha^{15.7}_{1.4} = -0.5$. The spectrum of this source is shown in Fig. \ref{fig:spectrum}, using data from Paper I; the source has an interesting spectrum as it is very steep ($\alpha_{0.61}^{1.4} = 1.5$) at lower frequencies, but then it turns up and rises at higher frequencies ($\alpha_{1.4}^{15.7} = -0.51$). The likely explanation for the rise at high frequencies is the presence of a compact core which dominates the emission at 15 GHz; however the emission at 1.4 GHz is likely still to be dominated by the more extended steep spectrum region with the core being too faint to be detected by the VLBA observations. This source highlights the complicated spectral shapes manifested at higher frequencies and the difficulties this presents when understanding and modelling this population.

\subsection{Source morphology}\label{section:morphology}

\begin{figure}
\centerline{\includegraphics[width=8cm]{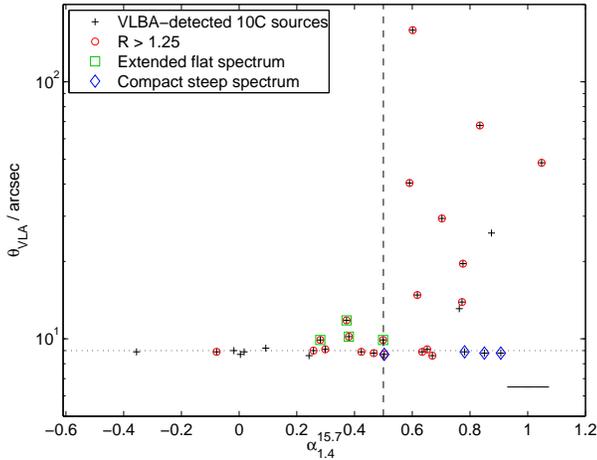}}

\caption{VLA size, $\theta_{\rm VLA}$, against spectral index, $\alpha^{15.7}_{1.4}$, for the 33 10C sources detected by the VLBA. Sources with $R > 1.25$ are circled in red (for comparison with Fig. \ref{fig:morph-alpha-10C}). The vertical dashed line is at $\alpha = 0.5$, the cutoff between steep and flat spectrum sources. The horizontal dotted line is at $\theta_{\rm VLA} = 9~\rm arcsec$, and sources with values of $\theta_{\rm VLA}$ larger than this are resolved. The four extended, flat spectrum sources are marked by squares and the compact, steep spectrum sources are marked by diamonds. For clarity individual error bars have been omitted but a point representing the median error in the x axis is shown in the bottom right hand corner for reference.}\label{fig:size-alpha-10C}
\end{figure}

\begin{figure}
\centerline{\includegraphics[width=8cm]{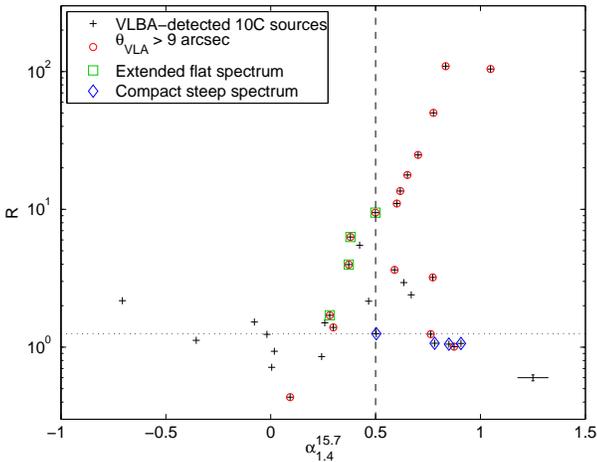}}

\caption{The ratio of the VLA integrated flux density to the VLBA integrated flux density ($R = S_{\rm VLA}/S_{\rm VLBA}$) against spectral index for the 33 10C sources detected by the VLBA. Sources with $\theta_{\rm VLA} > 9~\rm arcsec$ are circled in red (for comparison with Fig. \ref{fig:size-alpha-10C}). The vertical dashed line is at $\alpha = 0.5$, the cutoff between steep and flat spectrum sources, and the horizontal dotted line is at $R = 1.25$. The four extended, flat spectrum sources are marked by squares and the compact, steep spectrum sources are marked by diamonds.  For clarity individual error bars have been omitted but a point representing the median errors is shown in the bottom right hand corner for reference.}\label{fig:morph-alpha-10C}
\end{figure}

\begin{figure*}
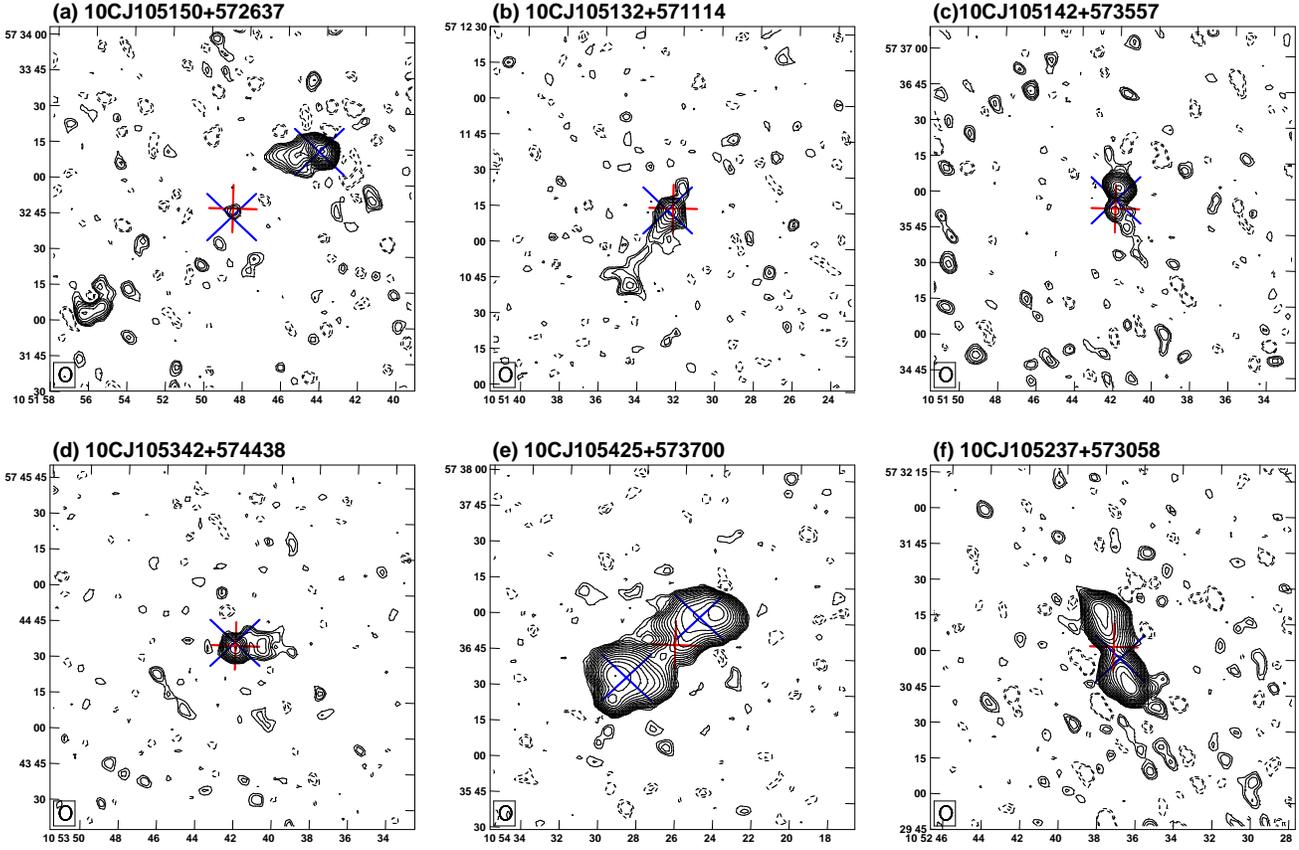

\centerline{\includegraphics[bb=60 214 557 700,clip=,width=5.5cm]{L0578.eps} \quad
            \includegraphics[bb=60 214 557 700,clip=,width=5.5cm]{L0477.eps} \quad
            \includegraphics[bb=60 214 557 700,clip=,width=5.5cm]{L0536.eps}}
								  
\bigskip							  
								  
\centerline{\includegraphics[bb=60 214 557 700,clip=,width=5.5cm]{L1235.eps} \quad
            \includegraphics[bb=60 214 557 700,clip=,width=5.5cm]{L1374.eps} \quad
            \includegraphics[bb=60 214 557 700,clip=,width=5.5cm]{L0890.eps}}
\caption{Example 610-MHz GMRT images of some extended sources detected by the VLBA (taken from \citep{2008MNRAS.387.1037G,2010BASI...38..103G} GMRT data). Source names are the IDs from the 10C catalogue, listed in Table \ref{tab:results_M10C}. Crosses ($\times$) show the positions of the 10C source, pluses ($+$) show the positions of the VLBA sources and are 20 arcsec across. The contours are drawn at $(\pm 2\sqrt{2^n}, n = 0, 1 \dots 7) \times a \textrm{ mJy}$ where $a$ = 0.062 for (a), 0.077 for (b), 0.061 for (c), 0.087 for (d), 0.098 for (e) and 0.077 for (f). \label{fig:contour-plots}}
\end{figure*}

%
%
\makeatletter
\let\@makecaption=\SFB@makefigurecaption
\makeatother
\setlength{\rotFPtop}{0pt plus 1fil}
\setlength{\rotFPbot}{0pt plus 1fil}

\begin{sidewaystable*}
\caption{Properties of sources detected by the VLBA which are not in the 10C catalogue. All flux densities listed are integrated flux densities. The two sources without any spectral index information are the two sources which lie very close to (but do not match to) a bright 10C source, meaning we are unable to place a limit on their flux density (see Section \ref{section:samples} for more details).}\label{tab:results_VD}
\small\rm
\begin{center}
\begin{tabular}{llldddddddddcdd}
\hline
 Name$^{\rm a}$ & RA (J2000) & Dec (J2000) & \dhead{$S_{\rm VLA}$} & \dhead{$\sigma\_S_{\rm VLA}$} & \dhead{$S_{\rm VLBA}$} & \dhead{$\sigma\_S_{\rm VLBA}$} & \dhead{$S_{\rm 10C}$ map$^{\rm b}$} & \dhead{$\sigma\_S_{\rm 10C}$ map$^{\rm b}$} & \dhead{10C 3$\sigma$ limit$^{\rm c}$} & \dhead{$\alpha^{15.7}_{1.4}$} & \dhead{$\sigma\_\alpha^{15.7}_{1.4}$} & Flag$^{\rm d}$ & \dhead{$\theta_{\rm {VLA}}$$^{\rm e}$} & \dhead{$R^{\rm f}$} \\
      &    &     & \dhead{/mJy}  & \dhead{/mJy}    & \dhead{/mJy} & \dhead{/mJy}    &  \dhead{/mJy} &  \dhead{/mJy}          & \dhead{/mJy}                 &         &            &      & \dhead{/arcsec}       &   \\
\hline
  J104922.9+571901 & 10:49:22.87 & +57:19:01.8 & 2.425 & 0.038 & 2.209 & 0.234  &       &       & 0.28 &  0.89  &      & 1 & 8.6 & 1.10\\
  J104951.3+572812 & 10:49:51.33 & +57:28:12.3 & 1.083 & 0.037 & 0.922 & 0.119  & 0.181 & 0.050 &      &  0.74  & 0.01 & 2 & 9.0 & 1.17\\
  J105012.6+571137 & 10:50:12.56 & +57:11:37.0 & 0.616 & 0.019 & 0.415 & 0.059  & 0.232 & 0.057 &      &  0.40  & 0.01 & 2 & 9.2 & 1.48\\
  J105025.5+570453 & 10:50:25.52 & +57:04:53.4 & 0.250 & 0.023 & 0.271 & 0.061  &       &       & 0.27 &  -0.04 &      & 1 & 8.3 & 0.92\\
  J105032.7+572646 & 10:50:32.67 & +57:26:46.6 & 0.466 & 0.015 & 0.497 & 0.064  & 0.194 & 0.035 &      &  0.36  & 0.01 & 2 & 8.7 & 0.94\\
  J105034.0+572152 & 10:50:34.03 & +57:21:52.4 & 1.005 & 0.017 & 0.220 & 0.042  &       &       & 0.12 &  0.87  &      & 1 & 8.9 & 4.57\\
  J105045.2+573734 & 10:50:45.23 & +57:37:34.0 & 0.858 & 0.034 & 0.290 & 0.081  &       &       & 0.15 &  0.74  &      & 1 & 9.1 & 2.96\\
  J105045.9+570822 & 10:50:45.90 & +57:08:22.4 & 0.344 & 0.017 & 0.153 & 0.042  &       &       & 0.19 &  0.25  &      & 1 & 8.9 & 2.25\\
  J105106.7+571152 & 10:51:06.71 & +57:11:52.8 & 0.355 & 0.017 & 0.282 & 0.043  & 0.468 & 0.150 &      &  -0.11 & 0.03 & 2 & 9.1 & 1.26\\
  J105117.6+571639 & 10:51:17.63 & +57:16:39.3 & 0.398 & 0.011 & 0.411 & 0.049  & 0.279 & 0.080 &      &  0.15  & 0.02 & 2 & 8.6 & 0.97\\
  J105120.8+573037 & 10:51:20.82 & +57:30:37.4 & 0.336 & 0.014 & 0.294 & 0.044  &       &       & 0.17 &  0.29  &      & 1 & 8.8 & 1.14\\
  J105120.9+573532 & 10:51:20.85 & +57:35:32.5 & 8.496 & 0.025 & 8.966 & 0.898  &       &       &      &        &      &   & 8.7 & 0.95\\
  J105134.1+570922 & 10:51:34.12 & +57:09:22.1 & 0.348 & 0.025 & 0.278 & 0.044  &       &       & 0.24 &  0.15  &      & 1 & 9.0 & 1.25\\
  J105158.9+572330 & 10:51:58.92 & +57:23:30.1 & 0.234 & 0.009 & 0.304 & 0.038  &       &       & 0.13 &  0.25  &      & 1 & 8.6 & 0.77\\
  J105207.5+571903 & 10:52:07.49 & +57:19:04.0 & 0.324 & 0.011 & 0.269 & 0.036  & 0.126 & 0.051 &      &  0.39  & 0.01 & 2 & 9.0 & 1.20\\
  J105211.0+572908 & 10:52:11.03 & +57:29:08.0 & 1.713 & 0.012 & 1.585 & 0.160  & 0.121 & 0.035 &      &  1.10  & 0.01 & 2 & 8.7 & 1.08\\
  J105230.6+571312 & 10:52:30.61 & +57:13:12.4 & 0.282 & 0.012 & 0.291 & 0.042  &       &       & 0.14 &  0.30  &      & 1 & 9.0 & 0.97\\
  J105231.8+570650 & 10:52:31.81 & +57:06:50.5 & 1.137 & 0.020 & 1.015 & 0.113  & 0.464 & 0.098 &      &  0.37  & 0.02 & 2 & 8.8 & 1.12\\
  J105234.0+573057 & 10:52:34.00 & +57:30:57.2 & 0.144 & 0.018 & 0.143 & 0.029  &       &       &      &        &      &   & 8.0 & 1.01\\
  J105245.3+573616 & 10:52:45.34 & +57:36:16.3 & 0.448 & 0.016 & 0.267 & 0.041  & 0.187 & 0.052 &      &  0.36  & 0.01 & 2 & 10.4 & 1.68\\
  J105250.0+574450 & 10:52:50.02 & +57:44:51.0 & 0.988 & 0.024 & 0.309 & 0.072  & 0.265 & 0.041 &      &  0.54  & 0.01 & 2 & 8.7 & 3.20\\
  J105258.0+570834 & 10:52:58.02 & +57:08:34.9 & 0.375 & 0.018 & 0.471 & 0.070  & 0.263 & 0.063 &      &  0.15  & 0.01 & 2 & 8.8 & 0.80\\
  J105304.5+571547 & 10:53:04.49 & +57:15:47.6 & 0.198 & 0.011 & 0.155 & 0.034  &       &       & 0.18 &  0.04  &      & 1 & 8.9 & 1.28\\
  J105308.1+572222 & 10:53:08.08 & +57:22:22.9 & 0.317 & 0.012 & 0.145 & 0.029  & 0.153 & 0.040 &      &  0.30  & 0.01 & 2 & 9.2 & 2.19\\
  J105314.0+572448 & 10:53:13.98 & +57:24:48.6 & 0.127 & 0.012 & 0.142 & 0.029  & 0.167 & 0.020 &      &  -0.11 & 0.01 & 2 & 9.8 & 0.89\\
  J105319.0+571851 & 10:53:19.02 & +57:18:51.8 & 0.548 & 0.014 & 0.229 & 0.037  & 0.156 & 0.048 &      &  0.52  & 0.01 & 2 & 9.1 & 2.39\\
  J105325.3+572911 & 10:53:25.31 & +57:29:11.8 & 0.602 & 0.012 & 0.199 & 0.033  &       &       & 0.14 &  0.61  &      & 1 & 9.2 & 3.03\\
  J105327.5+573316 & 10:53:27.49 & +57:33:17.1 & 0.195 & 0.013 & 0.219 & 0.036  &       &       & 0.13 &  0.18  &      & 1 & 8.6 & 0.89\\
  J105335.8+572157 & 10:53:35.85 & +57:21:57.3 & 0.147 & 0.010 & 0.226 & 0.038  &       &       & 0.12 &  0.07  &      & 1 & 8.4 & 0.65\\
  J105356.5+572244 & 10:53:56.46 & +57:22:44.8 & 0.249 & 0.014 & 0.258 & 0.044  &       &       & 0.14 &  0.24  &      & 1 & 8.8 & 0.97\\
  J105401.2+573207 & 10:54:01.21 & +57:32:07.5 & 0.267 & 0.013 & 0.142 & 0.039  &       &       & 0.12 &  0.32  &      & 1 & 8.6 & 1.88\\
  J105423.3+573446 & 10:54:23.32 & +57:34:46.4 & 0.661 & 0.020 & 0.598 & 0.078  &       &       & 0.31 &  0.32  &      & 1 & 8.5 & 1.11\\
\hline\end{tabular}

\end{center}
Notes:\\
a) Source name and positions from M2013 and \citeauthor{2009MNRAS.397..281I} catalogues.\\
b) 15.7-GHz flux density value from 10C map (see Section \ref{section:unmatched-VLBI} for details).\\
c) Three times the local r.m.s. noise in the 10C map.\\
d) The 15.7-GHz value used to calculate $\alpha^{15.7}_{1.4}$. Flag = 1, if upper limit is used, Flag = 2, if value from 10C map is used.\\
e) Maximum angular size taken from the \citet{2009MNRAS.397..281I} 1.4-GHz VLA catalogue. For single sources this value corresponds to twice the maximum Full Width Half Maximum (FWHM) of the synthesised beam (see Section \ref{section:morph-methods}).\\
f)  Ratio of the VLA integrated flux density to the VLBA integrated flux density (see Section \ref{section:morph-methods}).\\
%
%
\end{sidewaystable*}

The two measures used to probe source morphology on different scales are described in Section \ref{section:morph-methods}. In Paper I we investigated the extent of the radio emission of the 10C sources on arcsec scales using FIRST and GMRT data (beam size $\approx 5$~arcsec) along with WSRT data (beam size $\approx 11$~arcsec). These new VLBA data allow us to probe the structure of these sources on  much smaller scales ($\approx 10$~milliarcsec).

Figs. \ref{fig:size-alpha-10C} and \ref{fig:morph-alpha-10C} show these measures against spectral index for the 33 10C sources detected in the VLBA observations. Fig. \ref{fig:size-alpha-10C} shows that, as expected, all of the significantly extended sources (with $\theta_{\rm{VLA}}> 12$~arcsec) have a steep spectrum. There are, however, four sources which are resolved with the VLA ($9 < \theta_{\rm{VLA}} < 12$ arcsec) in the 10C sample which have flat spectra (these are marked with green squares in Figs. \ref{fig:size-alpha-10C} and \ref{fig:morph-alpha-10C}). This is consistent with the findings in Paper I that there is a small population of extended, flat spectrum sources present, indicating that emission from the cores of these extended sources is becoming dominant at higher frequencies.

Fig. \ref{fig:morph-alpha-10C} shows $R = S_{\rm VLA}/S_{\rm VLBA}$ as a function of spectral index. All sources with $\theta_{\rm{VLA}}> 9~\rm arcsec$ are circled for comparison with Fig. \ref{fig:size-alpha-10C}, this shows that unsurprisingly the majority of the sources with large values of $R$ have $\theta_{\rm{VLA}}> 9~\rm arcsec$ and are therefore resolved on arcsec scales. This plot also shows that most of the flat spectrum sources have smaller values of $R = S_{\rm VLA}/S_{\rm VLBA}$ than the steep spectrum sources, indicating that they are more dominated by emission from a compact core. All of the sources with rising spectra ($\alpha < 0$) have values of $R$ close to unity, indicating that all of the emission detected by the VLA is also detected by the VLBA observations and therefore, as expected, comes from a very compact region. There are six steep spectrum sources with values of $R$ close to unity ($ R < 1.25$). Two of these sources have values of $\theta_{\rm VLA} > 9~\rm arcsec$ therefore, although they may have a compact core, they are clearly extended on arcsec scales. Of the remaining four compact sources (which have $\theta_{\rm VLA} < 9~\rm arcsec$), one (10CJ105152+570950) has a steep spectrum down to 610~MHz ($\alpha^{15.7}_{1.4} = 0.85$, $\alpha^{1.4}_{0.61} = 0.70$), which is surprising for an object this compact. The lower frequency spectral index is slightly lower than the higher frequency spectral index so the spectrum may turn over below 610~MHz. One source has a spectrum which rises at lower frequencies (between 610~MHz and 1.4~GHz) so it therefore peaks between 1.4~GHz and 15.7~GHz, typical of a gigahertz peaked-spectrum (GPS) source, while the spectra of the other two sources peak at slightly lower frequencies, and are therefore more typical of compact, steep spectrum sources \citep{1998PASP..110..493O}. 

In Paper I we found in the faint 10C sample a small number of extended, flat spectrum sources, four of which are seen in this sample. Here we have also found that the faint 10C sample contains a number (4/33, 12 percent) of compact sources with steep spectra. This shows that although the majority of the steep spectrum sources are extended and the majority of the flat spectrum sources are compact, high frequency spectral index is not always a good indicator of source structure.

\citet{2013MNRAS.tmp.1815C} investigated the angular sizes of sources in the Australia Telecope 20 GHz (AT20G) catalogue, which has a flux density limit of 40~mJy so probes a population with higher flux densities than the 10C survey. \citeauthor{2013MNRAS.tmp.1815C} found that 77 percent of sources in the AT20G catalogue are compact, with angular sizes less than 0.15 arcsec. They produce a plot which is similar to Fig. \ref{fig:morph-alpha-10C}, in which most of the sources lie in approximately the same regions. They identify a population of compact steep-spectrum sources and find a smooth transition between this population and GPS sources.

\subsection{Extended source structure}\label{section:extended}

Further information about the structure of the extended sources can be gained by looking at GMRT images of these sources (from the \citealt{2008MNRAS.387.1037G,2010BASI...38..103G} data) which have a resolution of $6 \times 5~\rm arcsec^2$; GMRT images of six extended sources detected in the VLBA observations are shown in Fig. \ref{fig:contour-plots}. In their discussion of the properties of the VLBA-detected sources, M2013 assume that all the extended emission comes from star-formation. There is evidence here that this assumption is not true in a significant number of cases; the six sources shown in Fig. \ref{fig:contour-plots} display extended structures that appear to relate to the radio AGN, rather than star formation. Of the 33 10C sources detected by the VLBA, nine show evidence of extended structure relating to extragalatic radio activity. This is consistent with the idea discussed in Section \ref{section:detections} that this population is dominated by FRI sources, where the extended structure is due to emission produced by radio jets. M2013 use the assumption that all the extended emission originates from star formation to show that the most luminous AGN tend to have higher levels of star-formation activity; they conclude that this argues against quenching of star formation by AGN feedback and instead provides evidence that radio AGN activity might enhance star-formation. The results in this paper suggest that the underlying assumption is not correct in a significant number of cases and therefore these conclusions may not necessarily hold.

\subsection{Optical properties}

\citet{2012ApJS..198....1F} have calculated photometric redshifts for this part of the Lockman Hole field. Out of the 33 10C sources which are detected by the VLBA observations, 23 have photometric redshifts. The median redshift of these 10C sources is $0.91 \pm 0.16$, compared to a median redshift for all VLBA-detected sources of $0.91 \pm 0.14$. There is no significant difference in the source type classifications in the \citeauthor{2012ApJS..198....1F} catalogue of the VLBA-detected sources which are and are not in the 10C catalogue -- both samples are dominated by early-type or bulge dominated galaxies (along with heavily extinct starbursts, which is probably due to a degeneracy in the fit as discussed in M2013). A full analysis of the multi-wavelength properties of 10C sources in the whole Lockman Hole field will be presented in a future paper.

\section{Properties of VLBA sources}\label{section:results2}

\begin{figure}
\centerline{\includegraphics[width=8cm]{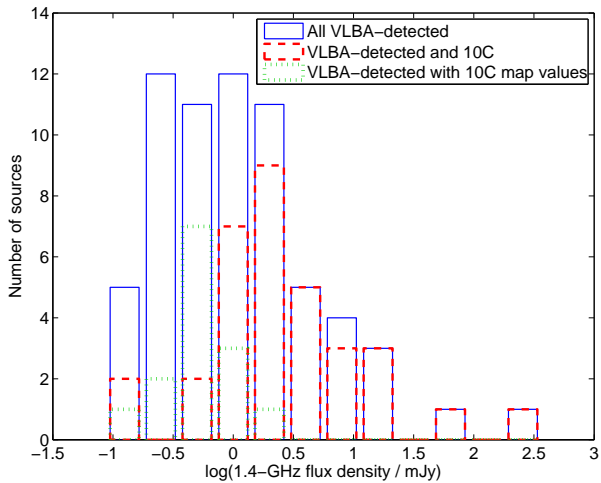}}
\caption{1.4-GHz flux density distribution for all 65 sources detected by the VLBA observations, with the subsample of sources also in the 10C sample shown separately, as are those sources which are not detected in the 10C catalogue for which a 15.7-GHz flux density value is available from the 10C map.}\label{fig:flux-VD}
\end{figure}

\begin{figure}
\centerline{\includegraphics[width=8cm]{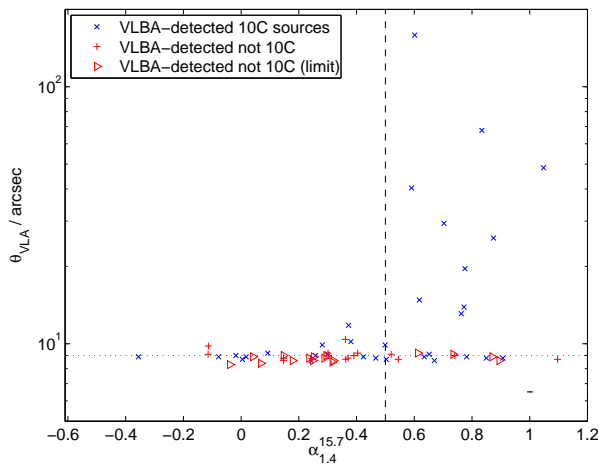}}

\caption{VLA size, $\theta_{\rm VLA}$, against spectral index, $\alpha^{15.7}_{1.4}$, for all sources detected by the VLBA observations. Sources which are also in the 10C sample are shown separately. Lower limits on spectral index are indicated by triangles. The vertical dashed line is at $\alpha = 0.5$, the cutoff between steep and flat spectrum sources. The horizontal dotted line is at $\theta_{\rm VLA} = 9~\rm arcsec$, and sources with values of $\theta_{\rm VLA}$ larger than this are resolved. For clarity individual error bars have been omitted but a point representing the median error in the x axis is shown in the bottom right hand corner for reference.}\label{fig:size-alpha}
\end{figure}

\begin{figure}
\centerline{\includegraphics[width=8cm]{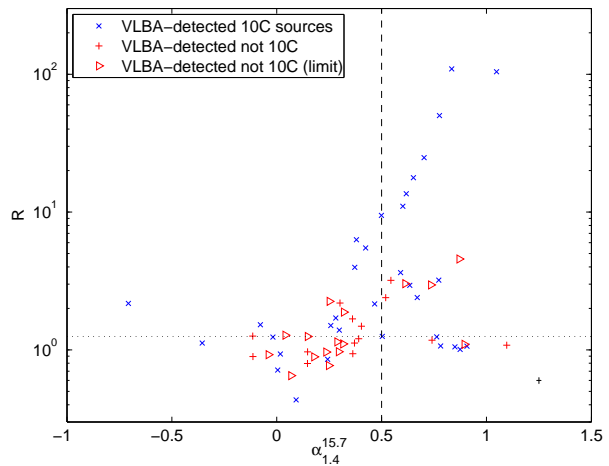}}

\caption{The ratio of the VLA integrated flux density to the VLBA integrated flux density ($R = S_{\rm VLA}/S_{\rm VLBA}$) against spectral index for all sources detected by the VLBA observation. Sources which are also in the 10C sample are shown separately. Lower limits on spectral index are indicated by triangles. The vertical dashed line is at $\alpha = 0.5$, the cutoff between steep and flat spectrum sources, and the horizontal dotted line is at $R=1.25$. For clarity individual error bars have been omitted but a point representing the median errors is shown in the bottom right hand corner for reference.}\label{fig:morph-alpha}
\end{figure}

In this section we look at the 15~GHz properties of the Middelberg VLBA-detected sample of 65 sources. The spectral index and morphology properties of the VLBA-detected sources which match to the 10C catalogue are given in Table \ref{tab:results_M10C}, the relevant rows being marked with an asterix. The properties of the 32 sources detected by the VLBA observations which are not in the 10C sample are summarised in Table \ref{tab:results_VD}. The 1.4-GHz flux density distributions of the VLBA-detected sources and those which are also in the 10C catalogue are shown in Fig. \ref{fig:flux-VD}, which shows that those sources which are in the 10C catalogue have larger 1.4-GHz flux densities than those which do not match to 10C. The sources which are not in the 10C catalogue for which a 15.7-GHz flux density value was calculated from the 10C map are also shown, these tend to be some of the brighter undetected sources. Fig. \ref{fig:size-alpha} shows the VLA size ($\theta_{\rm VLA}$) of all sources detected by the VLBA observations as a function of the spectral index $\alpha^{15.7}_{1.4}$. It indicates that \emph{all} of the significantly extended sources are detected by 10C, whereas those which are not detected at 15.7 GHz are not extended; this arises because the significantly extended sources in this sample have higher integrated flux densities and are therefore more likely also be in the 10C sample. Fig. \ref{fig:morph-alpha} shows $R$ against spectral index. The VLBA-detected sources which are not found in the 10C catalogue (shown by pluses and triangles) display a fairly uniform distribution of spectral indices, ranging from $\alpha = -0.11$ to 1.1. there are more flat spectrum sources than sources with steep spectra, but many of these values are lower limits so the spectra may be steeper in reality. There are at least three compact, steep spectrum sources (with $R < 1.25$) in addition to the four in the sample of 10C VLBA-detected sources, and there may be more as several of the other compact sources only have a lower limit on their spectral index.

\section{Conclusions}\label{section:conclusions}

Sixty-five percent (33/51) of the 10C sources in the VLBA survey area are detected by the VLBA observations, showing that these sources are AGN. The detected sources have a range of 15.7-GHz flux densities, with detected sources as faint as $S_{15.7~\rm{GHz}} = 0.3$~mJy. These results rule out the possibility discussed in Paper I that the 10C population is dominated by starforming or starbursting sources and provides strong evidence for our conclusion that the faint, high-frequency population is dominated by AGN, such as FRI sources. The proportions of sources in the $\rm S^3$ model are therefore probably roughly correct, i.e. the population at 1 mJy at 15~GHz is dominated by FRI sources. The spectral properties of these sources are not modelled correctly in the simulation, both because the flat spectrum cores are not correctly modelled, and because no spectral ageing has been included in the source model.

These results also show that there is a small but significant population (four out of 33) of very compact, steep spectrum sources in the 10C sample. They also show that in a number of cases the extended emission displayed by the sources in the VLBA sample is not caused by star formation as assumed by Middelberg et al., and is instead synchrotron emission produced by the AGN jets.

Multi-wavelength observations of these sources will be presented in a future paper and will provide further insights into the nature of this faint, high-frequency source population.

\section*{Acknowledgements}

IHW acknowledges an STFC studentship. We thank the anonymous referee for the thorough comments which have significantly improved this paper.

%
%

\setlength{\labelwidth}{0pt} 

\bsp

\label{lastpage}
\end{document}